\documentclass[trackchanges, twocolumn, twocolappendix, times]{aastex701}

\usepackage[T1]{fontenc}
\usepackage[utf8]{inputenc}
\DeclareUnicodeCharacter{2009}{\nobreakspace}
\usepackage[dvipsnames]{xcolor}
\usepackage{graphicx}
\usepackage{xspace}
\usepackage{amssymb, amsmath, mathtools}
\usepackage{bm}
\usepackage{microtype}
\usepackage[english]{babel}
\usepackage{hyperref}
\usepackage{times}
\usepackage[varg]{txfonts}

\newcommand{\norm}[1]{\left\lVert#1\right\rVert}

\newcommand{\td}{t_{\mathrm{d}}} 
\newcommand{\Ft}{\widetilde{\mathcal{F}}(t)} 
\newcommand{\mlz}{M_{\ell z}}
\newcommand{\ext}{{\mathrm{PMX}}} 
\newcommand{\fdm}{f_{\mathrm{DM}}} 
\newcommand{\dl}{D_{\ell}} 
\newcommand{\ds}{D_{s}} 
\newcommand{\dls}{D_{\ell s}} 
\newcommand{\paramsTrue}{\vec{\theta}^{\mathrm{tr}}} 
\newcommand{\hyp}[1]{\mathcal{H}_{#1}}

\newcommand{\ICTS}{\affiliation{International Centre for Theoretical Sciences, Tata Institute of Fundamental Research, Bengaluru 560089, India}}
\newcommand{\IUCAA}{\affiliation{The Inter-University Centre for Astronomy and Astrophysics (IUCAA), Post Bag 4, Ganeshkhind, Pune 411007, India}}
\newcommand{\NITAP}{\affiliation{Department of Physics, School of Sciences, National Institute of Technology Andhra Pradesh, Tadepalligudem 534101, India}}
\newcommand{\SHAO}{\affiliation{Shanghai Astronomical Observatory, Chinese Academy of Sciences, Shanghai 200030, People's Republic of China}}

\shorttitle{Host-Galaxy Effects on GW Microlensing Searches}
\shortauthors{U. Deka et al.}

\begin{document}

\title{Impact of the Host Galaxy on the Search for Gravitational-Wave Microlensing by Compact Objects}

\author[orcid=0000-0002-5942-4487]{Uddeepta Deka}
\email{uddeepta.deka@icts.res.in}
\ICTS
\SHAO

\author[orcid=0000-0001-7394-0755]{Apratim Ganguly}
\email{apratim@iucaa.in}
\IUCAA

\author[0000-0001-8753-7799]{Soham Bhattacharyya}
\email{xeonese@gmail.com}
\NITAP

\author[orcid=0000-0001-7519-2439]{Parameswaran Ajith}
\email{ajith@icts.res.in}
\ICTS

\hypersetup{pdfauthor={Deka et al.}}

\begin{abstract}
Gravitational-wave (GW) microlensing occurs when an intervening compact object causes diffraction of the GW, imprinting characteristic frequency-dependent modulations on the observed signal. Current searches for microlensing signatures in the detected GW events, including those used to constrain the fraction of dark matter in the form of compact objects, assume an isolated point mass lens (PML) model. However, astrophysical compact objects are typically embedded within larger structures, such as galaxies, whose potentials introduce additional convergence and tidal shear, thereby perturbing the lensed signal. In this work, we study the impact of the background galaxy in the search for microlensing by compact objects using a PML model. We distribute the compact objects inside a galaxy following the Navarro--Frenk--White density profile. We model the microlensing effects using the potential of a PML embedded in a constant convergence and shear, which are, in turn, calculated from a singular isothermal sphere model of the galaxy lens. We demonstrate that neglecting the galaxy's effects can systematically reduce sensitivity in microlensing searches, particularly for high microlens masses. Thus, future searches may need to incorporate the additional effect of the lens galaxy, particularly those aimed at placing robust constraints on the abundance of compact dark matter.
\end{abstract}



\section{Introduction}\label{sec:introduction}
Gravitational-wave (GW) observations from compact binary coalescences (CBCs) have enabled powerful probes of astrophysics~\citep{LIGOScientific:2017zic, LIGOScientific:2026ctl}, nuclear physics~\citep{LIGOScientific:2018cki}, cosmology~\citep{LIGOScientific:2026uyd}, and tests of general relativity~\citep{LIGOScientific:2026oim}. Since the first detection of GW150914~\citep{LIGOScientific:2016aoc}, nearly $400$ confident GW detections from CBC events have been reported by the LIGO--Virgo--KAGRA (LVK) collaboration~\citep{LIGOScientific:2014pky, VIRGO:2014yos, KAGRA:2020tym} until the second part of the fourth observing run~\citep{LIGOScientific:2026wfs}. Binary black hole (BBH) mergers constitute the bulk of the observed events; the remainder are associated with mergers of binary neutron stars and neutron star--black hole binaries. 

As GWs propagate from their source to the observer, they can encounter intervening compact objects (COs) and extended matter distributions, which can gravitationally lens the signal. Depending on the characteristic gravitational scale of the lens ($G M_\ell/c^2$), lensing manifests itself in two distinct regimes: when the gravitational scale of the lens is much larger than the GW wavelength ($GM_\ell/c^2 \gg\lambda_{\rm GW}$), lensing effects are accurately described by the geometric optics approximation. Galaxies, galaxy clusters, and sufficiently massive COs ($M_\ell\gtrsim10^5\, M_\odot$ in LVK observations) belong to this regime. If the source, lens, and observer are favourably aligned, multiple images are produced. While the angular separations between such images are typically too small to be measured using GW observations, their time delays could be measured well. These \textit{strongly lensed} signals share the same intrinsic frequency evolution but differ by magnification and by constant phase shifts~\citep{Dai:2017huk, Haris:2018vmn, Ezquiaga:2023xfe}. 

A qualitatively different situation arises when the wavelength becomes comparable to the gravitational scale of the lens ($G M_\ell /c^2 \sim \lambda_{\rm GW}$), where wave-optics effects become significant, and a single, frequency-modulated, diffracted signal is produced~\citep{Nakamura:1997sw, Takahashi:2003ix, Dai:2018enj}. We refer to the lensing of GWs in this regime as \textit{microlensing}. Unlike electromagnetic observations, where diffraction effects are effectively unobservable owing to their short wavelengths, GW observations provide a unique opportunity to probe wave-optics lensing directly. Microlensing of GWs in the LVK frequency band ($f_{\rm GW}\sim 10-10^3 \text{ Hz}$)~\citep{KAGRA:2013rdx} occurs for COs with masses in the range $M_\ell \sim 10^2 - 10^5 M_\odot$. Examples of such \emph{microlenses} include intermediate-mass black holes, primordial black holes (PBHs) and other exotic COs.

No confirmed detections of lensed GWs have been reported to date~\citep{Hannuksela:2019kle, LIGOScientific:2021izm, LIGOScientific:2023bwz, Janquart:2023mvf, LIGOScientific:2025cwb, Chakraborty:2025maj,Barsode:2026yqc}~\footnote{Although the massive BBH event GW231123 has been identified as an outlier in recent microlensing searches~\citep{Hu:2025lhv, Goyal:2025eqo, Shan:2025dcd, Chakraborty:2025pxt,Cheung:2026pky}, its lensing interpretation remains unconfirmed.}. This lack of confirmed lensing observations has enabled upper bounds on the fraction $\fdm$ of dark matter in the form of COs~\citep{Basak:2021ten, LIGOScientific:2023bwz, Barsode:2024wda}.

Searches for microlensing signatures in the LVK data are performed by computing the \emph{Bayes factor} between the microlensed and unlensed hypotheses. These searches, however, rely generally on the isolated point mass lens (PML) models, an assumption that neglects the galactic environments in which astrophysical microlenses reside. The potential of the galaxy may modify the local lens mapping, altering the time-delay surface and consequently the wave-optics diffraction pattern caused by the CO lens. In this work, we investigate the consequences of neglecting the effect of the background galaxy in the search for microlensing signatures caused by CO lenses. Specifically, we quantify the degradation in the Bayes factor due to the oversimplification of the lens model in a statistical sense. Finally, we discuss the implications of this for future GW microlensing searches and for constraints on the abundance of compact dark matter.

Previous studies have examined how additional microlensing by stars and black holes modifies strongly lensed GW signals and affects the identification of their multiple images~\citep{Diego2019, MeenaBagla2020, Mishra2021}. They showed that wave-optics distortions can become significant at sufficiently high macro-magnifications. \citet{Yeung:2021chy} further simulated microlenses embedded in a galaxy potential near moderately magnified macro images and investigated their detectability using isolated PML templates. Building on these studies, we quantify the loss of sensitivity of microlensing searches to a population of COs tracing the dark matter distribution of a galactic halo, without requiring the identification of multiple galaxy-scale images. We derive the corresponding distribution of external convergence and shear, then combine it with a high signal-to-noise ratio (SNR) approximation to estimate the population-averaged degradation in the Bayes factor when the galactic potential is neglected.

The remainder of this paper is arranged as follows: In Sec.~\ref{sec:lensing_diffraction}, we briefly review the formalism of GW lensing in the diffractive regime. In Sec.~\ref{sec:modified_PML}, we describe the lens model used in this study which incorporates the effects of an external galactic potential on a microlens. Sec.~\ref{sec:pop_PMX} details the generation of the microlens population within the galaxy, which is necessary for the statistical analysis. In Sec.~\ref{sec:results}, we present our results on the loss in search sensitivity due to neglecting the effect of the galaxy. Finally, we conclude in Sec.~\ref{sec:conclusion} with a discussion of our findings and their implications for future microlensing searches. Throughout the rest of this paper, we adopt geometrized units with $G=c=1$, where $G$ is Newton's gravitational constant, and $c$ is the speed of light in a vacuum.


\section{Lensing in the wave-optics regime}\label{sec:lensing_diffraction}

Gravitational lensing in the wave-optics regime modifies the GW signal through the complex, frequency-dependent magnification $\mathcal{F}(f; \vec{\Lambda})$, which encapsulates the amplitude and phase modulations induced by diffraction. Here, $\vec{\Lambda}$ denotes the complete set of parameters specifying the lensing configuration, including the lens parameters $\vec{\lambda}$ and the transverse source position $\vec{\eta}$ in the lens plane. Consequently, an unlensed frequency-domain waveform $h(f;\vec{\vartheta})$, characterized by the source parameters $\vec{\vartheta}$, is transformed into
\begin{equation}\label{eq:h_lensed}
    h_\ell(f; \vec{\vartheta}, \vec{\Lambda})= \mathcal{F}(f; \vec{\Lambda}) \times~ h(f; \vec{\vartheta}) ~,
\end{equation}
where $h_\ell(f; \vec{\vartheta}, \vec{\Lambda})$ is the lensed GW waveform.

Under the thin lens approximation\footnote{The (reasonable) approximation that the physical extent of the lens along the line of sight is much smaller than the observer--lens, lens--source, and observer--source distances.}, the magnification function can be computed either from the Kirchhoff diffraction integral~\citep{Schneider:1992bmb, Takahashi:2003ix} or equivalently from the path-integral formulation~\citep{Nakamura:1999uwi}, giving
\begin{equation}\label{eq:F_dimfull}
    \mathcal{F}(f;\vec{\Lambda})\equiv \frac{\ds}{\dl\dls}\frac{f}{i}(1+z_\ell)\int d^2\vec{\xi}\;\exp\left[i\, 2\pi f\, \td(\vec{\xi}; \vec{\Lambda})\right]~,
\end{equation}
where the normalization is chosen such that $|\mathcal{F}|=1$ in the absence of lensing. Here, $\dl$, $\ds$, and $\dls$ denote the angular diameter distances from the observer to the lens, the observer to the source, and the lens to the source, respectively, while $z_\ell$ is the lens redshift. The integration is performed over the lens plane, whose coordinates are denoted by the transverse vector $\vec{\xi}$, measured with respect to the optical axis~\footnote{The optical axis is the line joining the observer and the center of the lens.}. 

The observed waveform is obtained by coherently summing the contribution from all possible propagation paths through the lens plane, with the phase of each path determined by its corresponding time delay. Consequently, the diffraction pattern is entirely governed by the structure of the time-delay surface.
The time-delay function, $\td$, gives the excess travel time of a wave propagating from the source to the observer through the point $\vec{\xi}$ on the lens plane relative to the unlensed path. It is given by
\begin{equation}\label{eq:td_dimfull}
    \td (\vec{\xi}; \vec{\Lambda}) \equiv \frac{\dl\ds}{2\dls}\left(\frac{\vec{\xi}}{\dl}-\frac{\vec{\eta}}{\ds}\right)^2 - \hat{\psi}(\vec{\xi};\vec{\lambda})~.
\end{equation}
The first term is the geometric time delay arising from the additional path length traversed by the deflected wave front compared to the unlensed trajectory. The second term corresponds to the gravitational (or Shapiro) time delay and is described by the lensing potential $\hat{\psi}$. The explicit form of $\hat{\psi}$ depends on the underlying lens model and encapsulates the physical properties of the lens.

It is convenient to recast Equation~\eqref{eq:F_dimfull} in terms of dimensionless variables by adopting the Einstein radius of the lens as the characteristic length scale in the lens plane. To this end, we define
\begin{equation}
    \vec{x}\equiv\frac{\vec{\xi}}{\xi_0}, \qquad \vec{y}\equiv \frac{\dl}{\ds}\frac{\vec{\eta}}{\xi_0}~,
\end{equation}
where $\vec{x}$ and $\vec{y}$ denote the positions on the lens and source planes, respectively, measured in units of the Einstein radius, $\xi_0$
\begin{equation}\label{eq:einstein_radius}
    \xi_0\equiv\sqrt{4M_\ell\frac{\dl\dls}{\ds}}~.
\end{equation}
In addition, we also introduce the dimensionless frequency
\begin{equation}\label{eq:def_w}
    w \equiv \frac{\ds}{\dls\dl} \xi_0^2 (1+z_\ell) 2 \pi f = 8 \pi \mlz f~,
\end{equation}
where $\mlz\equiv M_\ell (1+z_\ell)$ is the redshifted lens mass. In terms of these quantities, lensing magnification Equation~\eqref{eq:F_dimfull} takes the form
\begin{equation}\label{eq:F_dimless}
    \mathcal{F}(w;\vec{\Lambda}) = \frac{w}{2\pi i}\int d^2\vec{x}\,\exp\left[i w\tau_\mathrm{d}(\vec{x};\vec{\Lambda})\right], 
\end{equation}
where
\begin{eqnarray}
\label{eq:td_dimless}
 \tau_\mathrm{d}(\vec{x};\vec{\Lambda}) &\equiv& \frac{\dl\dls}{\ds\xi_0^2} \, \frac{\td(\xi_0\vec{x};\vec{\Lambda})}{ (1+z_\ell)} =  \frac{\norm{\vec{x}-\vec{y}}^2}{2} - \psi(\vec{x};\vec{\lambda}), ~~ \mathrm{and}, \nonumber\\
\psi(\vec{x};\vec{\lambda}) &\equiv& \frac{\dl\dls}{\ds \xi_0^2} ~ \hat{\psi}(\xi_0\vec{x};\vec{\lambda}) ~ = \frac{1}{4M_\ell}\hat{\psi}(\xi_0\vec{x};\vec{\lambda})~,
\end{eqnarray}
define the dimensionless time-delay function and the corresponding dimensionless lensing potential, respectively.

Constructing lensed GW waveforms through Equation~\eqref{eq:h_lensed} ultimately requires evaluating the magnification function $\mathcal{F}(w;\vec{\Lambda})$ for a given lens model. Closed-form analytic expressions for $\mathcal{F}$ exist only for simple lenses such as the isolated PML~\citep{Takahashi:2003ix}. For more general lens models, the diffraction integral in Equation~\eqref{eq:F_dimless} must be evaluated numerically. Direct computation of this highly oscillatory integral is computationally demanding, although recently developed techniques have started addressing these challenges~\citep{Villarrubia-Rojo:2024xcj, Ephremidze:2026era}. 

In this work, we use the numerical method developed in~\citet{Deka:2024ecp}, which efficiently evaluates the inverse Fourier transform, $\Ft$,  of the frequency-domain magnification function for arbitrary lens potentials: 
\begin{equation}\label{eq:ft_def}
  \Ft \equiv \frac{1}{2\pi}\int_{-\infty}^{\infty}dw\frac{2\pi i}{w} \mathcal{F}(w) e^{-i w t} = \int d^2\vec{x}\,\delta(t-\tau_\mathrm{d}(\vec{x}))~,
\end{equation}
where, for brevity, the dependence on $\vec{\Lambda}$ has been suppressed. $\mathcal{F}(w)$ is then obtained through the corresponding Fourier transform.


\section{Microlens in a galactic potential}
\label{sec:modified_PML}

\begin{figure}[t]
  \centering
  \includegraphics[width=\columnwidth]{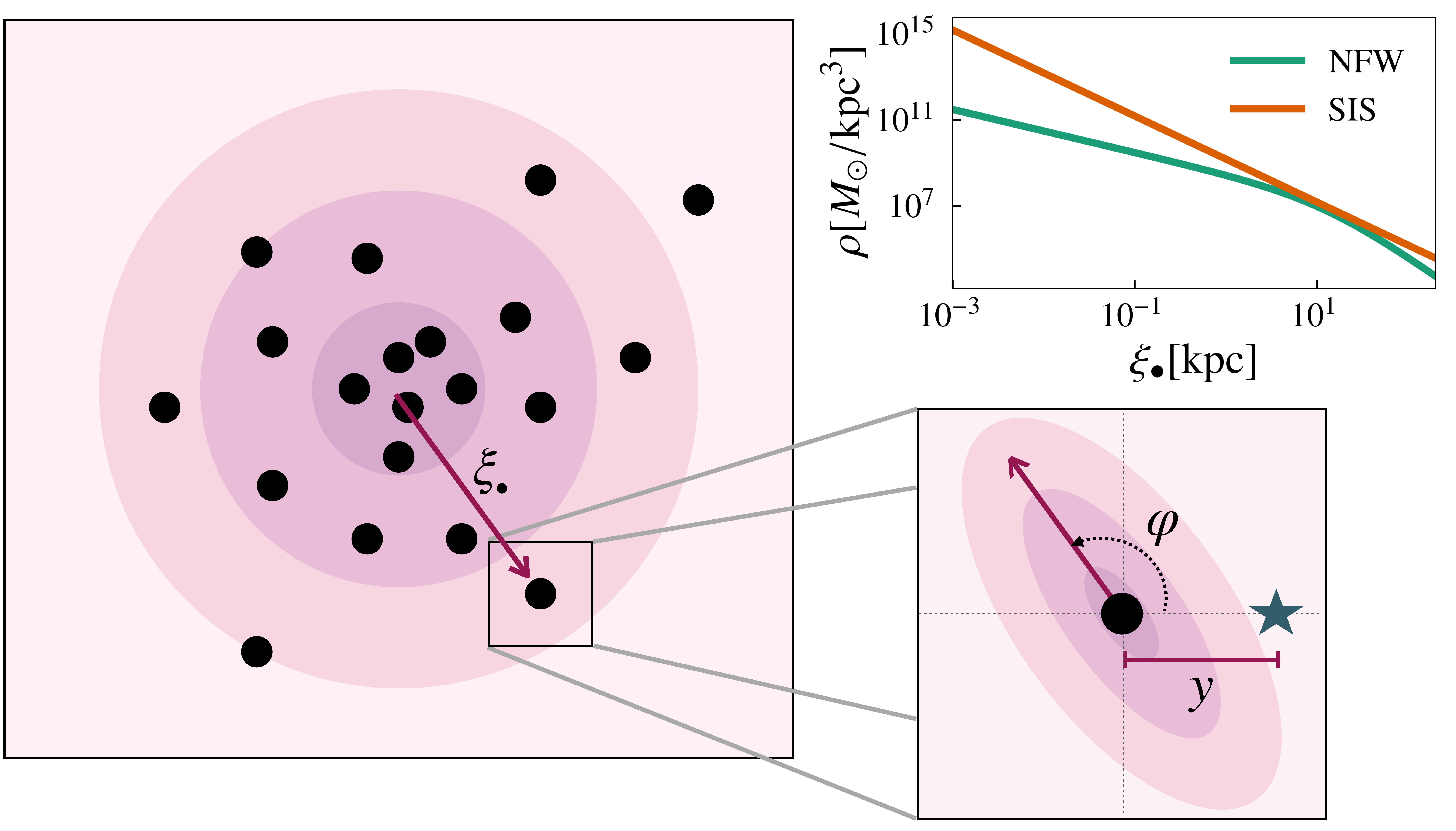}
  \caption{Schematic diagram of microlenses embedded in a galaxy. Left: microlenses (black circles) are located at a projected distance $\xi_\bullet$ from the galactic center. The highlighted region is enlarged in the lower-right panel, where the macrolens potential is approximated locally by constant convergence $\kappa$ and shear $\vec{\gamma}$. The angle $\varphi$ specifies the orientation of the shear, and $y$ denotes the dimensionless source position relative to the microlens. The upper-right panel compares the NFW and SIS density profiles. The lensing potential of the macro lens is modelled using an SIS profile, while the microlenses are distributed in the galaxy according to the NFW profile.}
  \label{fig:lensing_geometry}
\end{figure}

In this section, we describe the lens model used in this study. The simplest lens model is the PML, which describes stars and COs whose exterior spacetime is given by the Schwarzschild metric. In dimensionless coordinates introduced in the previous section, the lensing potential of the PML model is
\begin{equation}\label{eq:lensing_potential_PML}
  \psi_{\rm PML} = \ln (x)~,
\end{equation}
where $x\equiv \norm{\vec{x}}$ is the radial distance on the lens plane, assuming that the optical axis passes through the microlens.

In realistic astrophysical scenarios, however, microlenses (e.g., black holes) are seldom isolated. Instead, they are often embedded within a larger-scale mass distribution, such as a galaxy, which deforms the lensing potential by the PML. To account for these external effects, the lensing potential is augmented by including external \emph{shear} and \emph{convergence}. The shear $\vec{\gamma}=\{\gamma_1, \gamma_2\}$ characterizes the quadrupolar tidal gravitational field due to the galaxy, while the convergence $\kappa$ reflects the local isotropic modification to the projected mass density.

Although both $\kappa$ and $\vec{\gamma}$ depend on the location of the microlens as measured from the galactic center, within the Fresnel radius of the microlens, both of these parameters are effectively constant (see Appendix~\ref{app:model_validity} for details). We refer to this \emph{point mass in a constant external potential} as the PMX model, with the lensing potential given by~\citep{microlensing-book,Saha_2010}, 
\begin{equation}\label{eq:lensing_potential_PMX}
  \psi_\ext = \psi_{\rm PML} + \frac{\kappa}{2}\left(x_1^2 + x_2^2\right) + \frac{\gamma_1}{2}\left(x_1^2 - x_2^2\right) + \gamma_2 x_1 x_2~.
\end{equation}

Note that the origin of the coordinate system is attached to the microlens. As mentioned earlier, we consider the microlenses to lie in a galaxy. Figure~\ref{fig:lensing_geometry} illustrates this scenario. Throughout this work, we model the host galaxy as described by a singular isothermal sphere (SIS) profile. Note that this profile models the total matter density in the galaxy, including dark and baryonic matter. In this case, the convergence and shear at the location of the microlens satisfy~\citep{kormann1994},
\begin{subequations}\label{eq:kappa_SIS}
  \begin{align}
  \kappa &= \frac{1}{2\xi_\bullet}\frac{4\pi \sigma^2\dl\dls}{\ds}~,\label{eq:kappa_SIS_a}\\
  \gamma_1 &= -\kappa\cos 2\varphi~,\label{eq:kappa_SIS_b}\\
  \gamma_2 &= -\kappa\sin 2\varphi~\label{eq:kappa_SIS_c},
  \end{align}
\end{subequations}
where $\xi_\bullet$ denotes the projected distance between the microlens and the galactic center, $\sigma$ is the velocity dispersion of the galaxy lens (SIS), and $\varphi$ specifies the orientation of the external shear. Thus, the external potential is solely parameterised by $\kappa$ and $\varphi$. Finally, without loss of generality, we choose the source location as $\vec{y} = \{y, 0\}$, since an arbitrary source orientation can be absorbed into a corresponding rotation of the external shear.

\begin{figure*}[t]
    \centering
    \includegraphics[width=\textwidth]{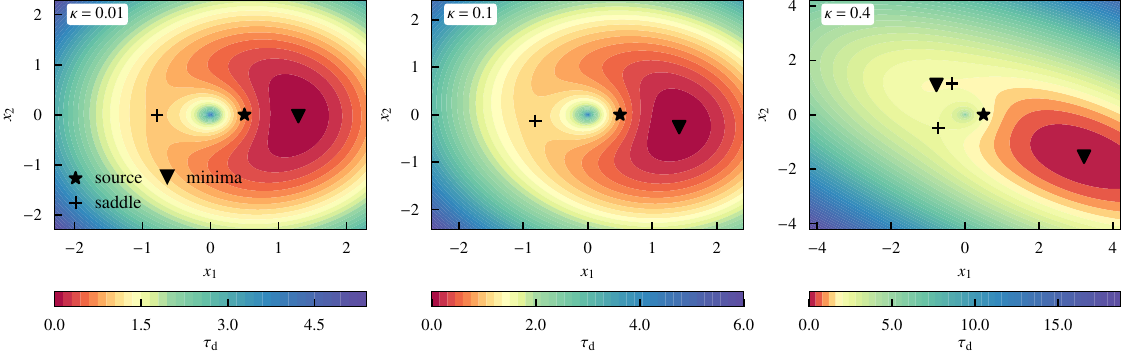}
    \caption{Contours of the {dimensionless} time delay, $\tau_\mathrm{d}$, on the lens plane coordinates ($x_1, x_2$). The microlens is located at the origin, while the source position $(y=0.5)$ is marked by a star. The panels from left to right correspond to $\kappa=0.01, 0.1$ and $0.4$, respectively. For each of these cases, the locations of the images and their types are also indicated. The shear angle is kept fixed at $\varphi=\pi/3$.}
    \label{fig:td_map}
\end{figure*}

Figure~\ref{fig:td_map} illustrates the contours of the time-delay function $\tau_\mathrm{d}$ on the lens plane for a fixed source position $y=0.5$ and shear angle $\varphi=\pi/3$. The panels from left to right correspond to increasing values of $\kappa=0.01, 0.1$ and $0.4$, respectively. The stationary points of $\tau_\mathrm{d}$, which correspond to the lensed images, are also indicated together with their Morse types (image parity). As $\kappa$ increases, the topology of the $\tau_\mathrm{d}$ surface changes, leading to different image configurations, reflecting the influence of the external potential. A similar effect can be observed upon varying $\varphi$ and $y$.

\begin{figure*}[t]
    \centering
    \includegraphics[width=\textwidth]{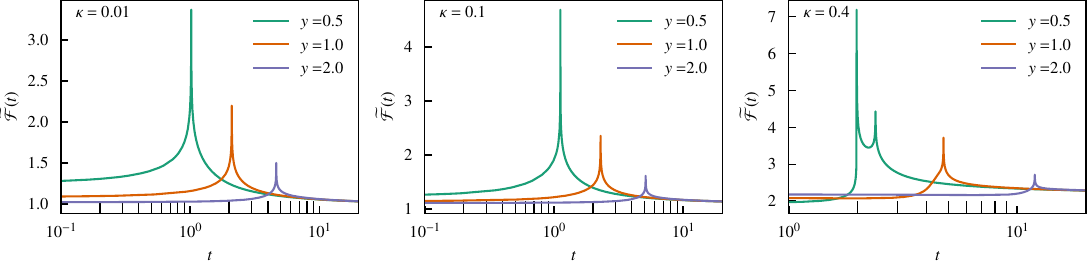}
    \caption{The time-domain magnification function $\Ft$ as a function of time $t$ (in units of $4\mlz$, where the global minimum lies at $t=0$). The panels from left to right correspond to cases with $\kappa=0.01, 0.1$ and $0.4$, respectively, for a fixed $\varphi = \pi/3$. The colors in each of these panels correspond to various source locations $y=0.5, 1.0$ and $2.0$. In the left and center panels, there are always two images --- a minimum image (at $t=0$) and a saddle image (identified by the logarithmic peak). In the right panel, for $y=0.5$, there are two saddle images and two minima images.}
    \label{fig:Ft_plots}
\end{figure*}

Figure~\ref{fig:Ft_plots} shows the time-domain magnification function $\Ft$ evaluated for different values of $\kappa$ and $y$, with the shear orientation fixed at $\varphi=\pi/3$. The global minimum image always arrives first and is therefore located at $t=0$. The contributions from the saddle images manifest as logarithmic peaks in $\Ft$, whereas minima and maxima produce step discontinuities. For example, when $y=0.5$, the left ($\kappa=0.01$) and middle ($\kappa=0.1$) panels exhibit the characteristic signatures of one minimum (at $t=0$) and one saddle image, while the right panel ($\kappa=0.4$) contains two minima and two saddle images. In the latter case, the second minimum image has a time delay very similar to that of one of the saddle images, making it difficult to visually identify from the $\Ft$ plot. These image configurations are supported by the corresponding time-delay surfaces shown in Figure~\ref{fig:td_map}. 

Although the singular features of $\Ft$ are determined by the time delays and parity of the images, Equation~\eqref{eq:ft_def} shows that $\Ft$ itself is obtained by integrating over the entire lens plane rather than by summing over the individual image contributions. The remaining plots in Figure~\ref{fig:Ft_plots} similarly showcase the expected $\Ft$ in the parameter space, which results in the two-image configurations. Due to the external potential, as $t\to\infty$, $\Ft\to 1/\sqrt{1-2\kappa}$.

\begin{figure*}[t]
    \centering
    \includegraphics[width=\textwidth]{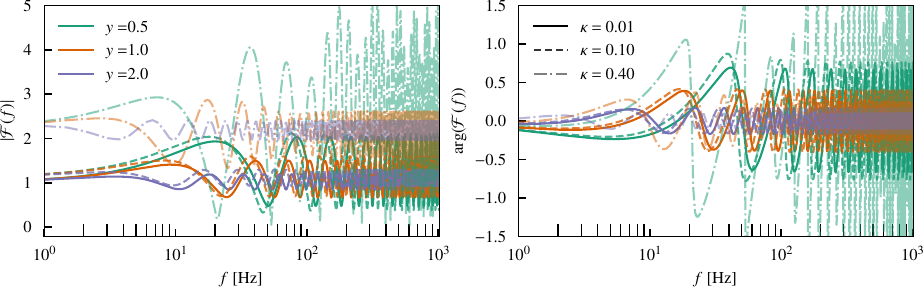}
    \caption{Amplitude (left) and phase (right) of the frequency-domain magnification function $\mathcal{F}(f)$ as a function of the GW frequency for a microlens of mass $M_\ell = 500 M_\odot$, located at redshift $z_\ell = 0.5$. The plots correspond to $\kappa=0.01$ (solid), $\kappa=0.1$ (dashed) and $\kappa=0.4$ (dashed--dotted). The shear angle is kept fixed at $\varphi=\pi/3$. Different colors indicate source positions $y=0.5, 1.0$ and $2.0$.}
    \label{fig:Ff_plots}
\end{figure*}

The corresponding frequency-domain magnification functions $\mathcal{F}(f)$ for the same lensing configurations considered in Figure~\ref{fig:Ft_plots} are shown in Figure~\ref{fig:Ff_plots}. The left and right panels display the amplitude and phase of $\mathcal{F}(f)$, respectively. As expected, the microlensing-induced modulations become increasingly pronounced for smaller source positions $y$ and larger values of convergence $\kappa$. Moreover, a nonzero $\kappa$ introduces an overall magnification offset, which increases with increasing $\kappa$. To express in physical frequency $f$, we adopt a microlens mass of $M_\ell = 500 M_\odot$ located at a redshift $z_\ell=0.5$.

\begin{figure*}[t]
    \centering
    \includegraphics[width=\textwidth]{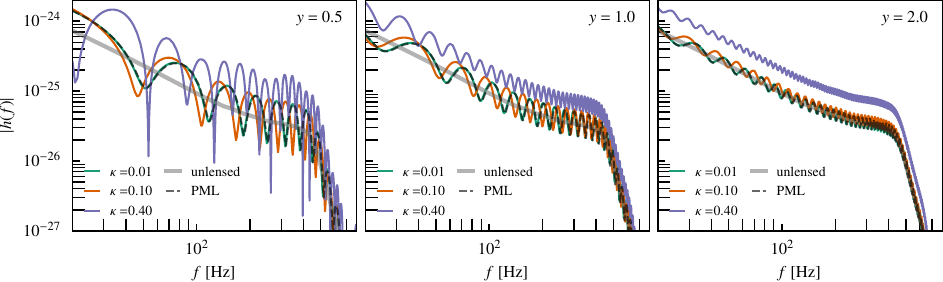}
    \caption{Amplitude of the frequency-domain GW signals from a nonspinning, equal-mass binary with component masses $20 M_\odot$, located at redshift $z_s = 2$. Gray solid curves show the unlensed signal, while the black dashed curves represent microlensed waveforms due to a PML of mass $500 M_\odot$ at redshift $z_\ell = 0.5$. Colored curves correspond to microlensing by PMX lenses with $\kappa=0.01, 0.1$ and $0.4$. The shear angle is kept fixed at $\varphi=\pi/3$. Panels from left to right correspond to source positions $y=0.5, 1.0$ and $2.0$, respectively.}
    \label{fig:hf_plots}
\end{figure*}

Finally, the lensed GW waveforms can be obtained using Equation~\eqref{eq:h_lensed}. Figure~\ref{fig:hf_plots} shows the amplitude of the resulting waveforms for a nonspinning, equal-mass binary with component masses of $20 M_\odot$ at a source redshift of $z_s = 2$. The gray solid curves represent the unlensed waveform, while the black dashed curves correspond to microlensed waveforms by an isolated PML with $M_\ell=500 M_\odot$ at $z_\ell = 0.5$. The colored curves show the corresponding microlensed waveforms for PMX lens configurations shown in Figure~\ref{fig:Ff_plots}. Across all panels, the microlensed waveforms deviate noticeably from the unlensed case. The deviation from the PML model is also prominent for large values of $\kappa$. As the source location $y$ increases, the strength of microlensing decreases as expected. At the same time, the introduction of an overall magnification offset due to a nonzero $\kappa$ raises the baseline amplitude in each plot.

\section{Populating microlenses in a galactic potential}\label{sec:pop_PMX}

To evaluate the loss in sensitivity of a search for microlensing in a statistical sense, we consider a population of COs distributed within a galaxy. Compact dark matter candidates, such as PBHs, are expected to trace the underlying dark matter distribution of their host halos. Under this assumption, the spatial distribution of PBH microlenses follows that of the galactic dark matter halo. 

In this work, we model the host halo as a Milky Way--like galaxy described by the Navarro--Frenk--White (NFW) density profile~\citep{Navarro:1996gj}. The NFW profile emerges naturally from cosmological $N$-body simulations of hierarchical structure formation~\citep{Bullock:1999he} and provides an accurate description of the density profiles of virialized dark matter halos across a broad range of mass scales. GW microlensing has been used to constrain the fraction of dark matter in the form of COs such as PBHs~\citep{Basak:2021ten, LIGOScientific:2023bwz}. In this scenario, PBH microlenses will be distributed according to the dark matter profile. The projected NFW profile therefore determines the probability of finding a PBH microlens at a projected radial distance between $\xi_\bullet$ and $\xi_\bullet+d\xi_\bullet$ from the halo center on the lens plane,
\begin{equation}\label{eq:microlens_spatial_dist}
  d p \propto \Sigma_{\rm NFW}(\xi_\bullet) \, 2\pi \xi_\bullet\,d\xi_\bullet~,
\end{equation}
where $\Sigma_{\rm NFW}$ is the surface mass density of the NFW halo given by~\citep{Bartelmann:1996hq},
\begin{equation}\label{eq:sigma_NFW}
  \Sigma_{\rm NFW}(\zeta) = 2\,\rho_0\,a_0\frac{1-\mathcal{I}(\zeta)}{\zeta^2-1}~,
\end{equation}
where $\zeta\equiv\xi_\bullet/a_0$ is the dimensionless projected radius, $\rho_0$ is the characteristic density, and $a_0$ is the characteristic scale radius of the halo. Here, the NFW profile is used solely to sample PBH positions since they should follow the dark matter distribution; the SIS model determines the total lensing effect of the host galaxy, which is determined by the dark matter and baryonic matter. For the Milky Way halo, we consider $\rho_0\sim 10^7 M_\odot\mathrm{kpc}^{-3}$ and $a_0\sim 17 \mathrm{kpc}$~\citep{Nesti:2013uwa}. The function $\mathcal{I}(\zeta)$ is defined as
\begin{equation}
  \mathcal{I}(\zeta) =
  \begin{cases}
    \frac{1}{\sqrt{\zeta^2-1}}\tan^{-1}\sqrt{\zeta^2-1}, & \zeta >1~;\\
    \frac{1}{\sqrt{1-\zeta^2}}\tanh^{-1}\sqrt{1-\zeta^2}, & \zeta <1~;\\
    1, & \zeta =1~.
  \end{cases}
\end{equation}

\begin{figure}[htbp]
  \centering
  \includegraphics[width=0.473\textwidth]{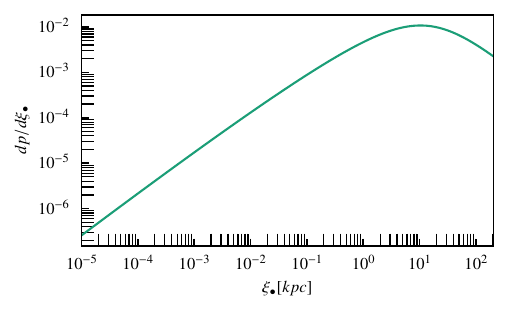}
  \caption{Expected distribution of the location of individual microlenses measured from the center of the galaxy halo, assuming that they follow the dark matter distribution.}
  \label{fig:microlens_spatial_dist}
\end{figure}

Figure~\ref{fig:microlens_spatial_dist} shows the resulting distribution of microlens positions measured from the halo center. The distribution peaks at a projected distance of $\xi_\bullet \sim 0.5 a_0$ and falls off at larger radii. The distribution is truncated at the virial radius of the halo, $r_{\rm vir}\sim 200~\mathrm{kpc}$, beyond which the density of the halo drops sharply.

\begin{figure}[htbp]
  \centering
  \includegraphics{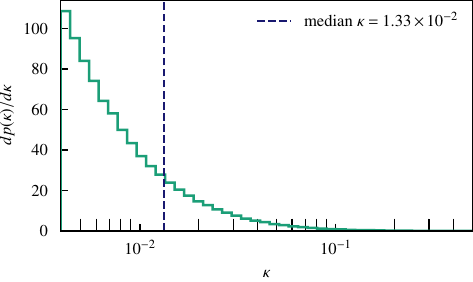}
  \caption{Distribution of $\kappa$ at the microlens locations for an SIS galaxy with $\sigma=200 \mathrm{ km/s}$, $\dl=500 \mathrm{ Mpc}$, and  $\ds=1000 \mathrm{ Mpc}$.}
  \label{fig:kappa_SIS_dist}
\end{figure}

Given the spatial distribution of microlenses in Equation~\eqref{eq:microlens_spatial_dist}, the distribution of external $\kappa$ experienced by these microlenses can be obtained using Equation~\eqref{eq:kappa_SIS_a}. Assuming a velocity dispersion $\sigma=200\mathrm{km/s}$ and a fiducial lensing geometry with $\dl=500\, \mathrm{Mpc}$ and $\ds=1000 \, \mathrm{Mpc}$, we obtain the distribution shown in Figure~\ref{fig:kappa_SIS_dist}. Note that microlenses residing deep in the galaxy's potential ($\zeta \ll 1$) produce large values of $\kappa$ ($\kappa \sim 0.1$). However, there will be many more microlenses with $\zeta > 1$, producing small values of $\kappa$ ($\kappa < 0.01$). Thus, the distribution is peaked at low values of $\kappa$, with a median value of $\kappa\simeq0.013$.

Finally, we assume that the microlenses are isotropically distributed in azimuthal angle, such that the shear angle $\varphi$ is uniformly distributed between $0$ and $2\pi$. We draw the source positions uniformly in area, i.e., $p(y) \propto y$, with $0.1\leq y\leq 3$. These ingredients fully specify the ensemble of PMX lens configurations that we use to evaluate the loss in microlensing significance in the next section.


\section{Loss in the search sensitivity}\label{sec:results}

Given observational data $d$ that contain a GW signal, current searches evaluate the evidence for microlensing by computing the Bayes factor between the two competing hypotheses: the microlensed (PML-lensed) hypothesis $\mathcal{H}_{\mathrm{PML}}$, and the unlensed hypothesis $\mathcal{H}_{\mathrm{U}}$. The Bayes factor is defined as
\begin{equation}
\mathcal{B}_\mathrm{U}^\mathrm{PML}\equiv\frac{p(d|\mathcal{H}_{\mathrm{PML}})}{p(d|\mathcal{H}_{\mathrm{U}})}~,
\end{equation}
where $p(d|\mathcal{H}_{\mathrm{PML}})$ and $p(d|\mathcal{H}_{\mathrm{U}})$ are the Bayesian evidences (i.e., marginalized likelihoods) under the PML-lensed and unlensed hypotheses, respectively. A Bayes factor $\mathcal{B}_\mathrm{U}^\mathrm{PML}>1$ indicates that the data favor the microlensed scenario over the unlensed one, while values less than unity suggest greater support for the unlensed hypothesis. Current GW microlensing searches employ this Bayes factor as a statistical discriminator to infer the presence of microlensing signatures.

Similarly, we can define the Bayes factor $\mathcal{B}_\mathrm{U}^\mathrm{PMX}$ to compare the PMX-lensed and unlensed hypotheses, thereby enabling the assessment of microlensing signatures when host galaxy effects on the microlens are included:

\begin{equation}
    \mathcal{B}_\mathrm{U}^\mathrm{PMX}\equiv\frac{p(d|\mathcal{H}_{\mathrm{PMX}})}{p(d|\mathcal{H}_{\mathrm{U}})}~.
\end{equation}

Suppose the true signal is microlensed by a PMX lens with parameters $\vec\theta^{\mathrm{tr}}$ (including both source as well as lensing parameters $M_{\ell z}, y, \kappa,$ and $\varphi$), but the inference is carried out using the simplified PML model and the corresponding Bayes factor $\mathcal{B}_\mathrm{U}^\mathrm{PML}$. We aim to investigate whether neglecting the influence of the host galaxy, that is, using an incorrect lensing model, significantly degrades the Bayes factor, hence reducing the sensitivity to microlensing.

To this end, we define the \emph{retention factor} as the ratio between the ``suboptimal'' (using the PML model) and ``optimal'' (using the PMX model)  Bayes factors,
\begin{equation}\label{eq:PMX-loss-factor}
    \mathcal{R}_B \equiv \frac{\mathcal{B}_\mathrm{U}^\mathrm{PML}}{\mathcal{B}_\mathrm{U}^\mathrm{PMX}} \simeq \exp\left[-\frac{\rho^2}{2}(1 - \mathrm{FF}^2)\right]~,
\end{equation}
where the last equality holds only in the high-SNR ($\rho$) limit (see Appendix~\ref{app:loss_factor_derivaiton} for a derivation). The \emph{fitting factor} ($\mathrm{FF}$) is defined as
\begin{equation}\label{eq:PMX-fitting-factor}
    \mathrm{FF}\equiv \max_{\vec\theta_\mathrm{PML}}\mathcal{M}(\vec{\theta}^{\mathrm{tr}}, \vec{\theta}_{\mathrm{PML}})~,
\end{equation}
where the match $\mathcal{M}$ is computed between the true PMX-lensed waveform and the best-fit waveform from the PML template family. The maximization of the match is performed over the full parameter space $\vec\theta_\mathrm{PML}$, encompassing both lens and intrinsic source parameters.

We compute the retention factor following the steps outlined below:
\begin{enumerate}
    \item \emph{Generate lensed injection signals}. We simulate a grid of PMX-lensed injection (or true) signals spanning $\kappa\in[10^{-3},10^{-1}]$, $y\in [0.1, 3.0]$ and $M_{\ell z}\in[10^2, 10^5]M_\odot$. The shear angle $\varphi$ is drawn uniformly from $[0, 2\pi)$. The source parameters include chirp mass $\mathcal{M}_c=\{20, 30, 40\}M_\odot$ and symmetric mass ratio $\eta=\{0.2, 0.25\}$, with all systems assumed to be nonspinning, that is, effective spin $\chi_\mathrm{eff}= 0$ \citep{Ajith:2009bn} and $\chi_\mathrm{p}=0$~\citep{Hannam:2013oca}. These quantities define the true parameter vector $\vec\theta^\mathrm{tr}$ for each simulated signal. Throughout this study, the GW signal is modeled using the \textsc{IMRPhenomXPHM} waveform family~\citep{Pratten:2020ceb}.
    \item \emph{Fitting factor computation}. For each of these true signals, we compute the FF by maximizing the match over the PML parameter space, comprising $M_{\ell z}, y, \mathcal{M}_c, \eta, \chi_\mathrm{eff}$ and $\chi_\mathrm{p}$. The matches are evaluated using the \textsc{PyCBC} software package~\citep{alex_nitz_2024_10473621}. We use the noise power spectral density (PSD) pertaining to the advanced LIGO targeted for the O5 observing run (A+ configuration)~\citep{LIGOScientific:2014pky}.
    \item \emph{Retention factor estimation}. For a given SNR, inserting $\mathrm{FF}$ into Equation~\eqref{eq:PMX-loss-factor} gives the corresponding retention factor.
    \item \emph{Reweighting and marginalization}. The computed retention factors are reweighted and marginalized over the true lensing parameters ($\kappa^\mathrm{tr}, y^{\mathrm{tr}}$) using their probability distributions as
    \begin{equation}\label{eq:PMX-average-ff}
        \left\langle\frac{\mathcal{B}_\mathrm{U}^\mathrm{PML}}{\mathcal{B}_\mathrm{U}^\mathrm{PMX}}\right\rangle =\iiint d\varphi^{\mathrm{tr}} \,d\kappa^{\mathrm{tr}}\,dy^\mathrm{tr}\,p(\varphi^{\mathrm{tr}})\,p(\kappa^\mathrm{tr}) \,p(y^\mathrm{tr})\, 
        \mathcal{R}_B. 
    \end{equation}
    This gives the average retention factor over the microlens population.
\end{enumerate}

\begin{figure*}[htbp]
    \centering
    \includegraphics[width=\textwidth]{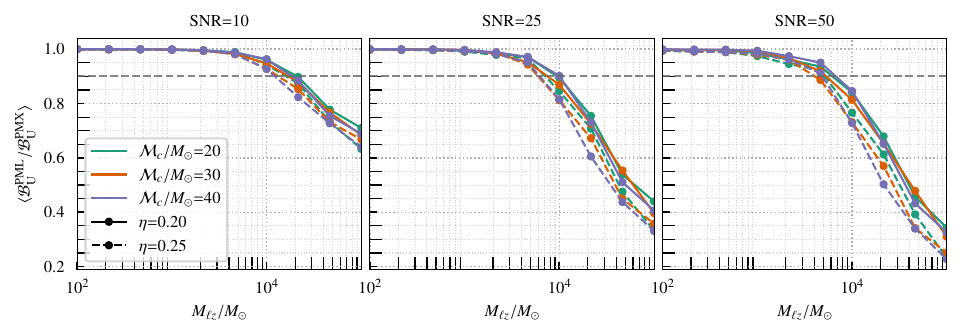}
    \caption{Retention factor as a function of the redshifted microlens mass. Each panel corresponds to a different SNR value (10, 25, and 50 from left to right). Colors distinguish binary chirp masses, $\mathcal{M}_c$, while line styles distinguish symmetric mass ratios, $\eta$.}
    \label{fig:PMX-blu_ratio}
\end{figure*}

Figure~\ref{fig:PMX-blu_ratio} presents the average retention factor pertaining to the microlens population as a function of the redshifted microlens mass for different values of chirp mass $\mathcal{M}_c$ and symmetric mass ratio $\eta$ for the sources. The SNR increases from the left to right panels. The plots show that the degradation in the Bayes factor becomes significant for higher microlens masses, whereas the simplified PML model remains an adequate approximation at lower lens masses. As expected from Equation~\eqref{eq:PMX-loss-factor}, the fractional loss is larger for events with higher SNRs.


\section{Conclusion}\label{sec:conclusion}

In this work, we investigated the effect of the lensing potential of the host galaxy on the search for microlensing by COs using a PML model. The additional effect of the galaxy lens was modeled locally as a constant convergence and shear, added to the microlens potential (PMX model). To assess the effect of using an approximate model for the microlens (the PML model, which neglects the additional contribution by the galaxy lens), we defined the retention factor --- the ratio of the lensing Bayes factors computed using the PMX and PML lensing models. In the high-SNR limit, this quantity depends on the mismatch between the true and best-fit waveforms, as well as the optimal SNR of the lensed signal.

Our analysis involved simulating a set of PMX-lensed GW signals over a broad parameter space of lens and source properties. We computed the FF between these signals and their best-fitting counterparts from the PML model. Lower values of the FF reflect the inability of the PML model to capture the additional waveform distortions introduced by the host galaxy's gravitational field.

We work under the hypothesis that these microlenses (such as PBHs) are dark matter candidates and are distributed in the galaxy following the dark matter distribution (NFW profile). We obtain the corresponding distributions of the convergence and shear that will be produced by the host galaxy in the locations of these microlenses (assuming an SIS profile for the galaxy lens). By marginalizing the retention factors over the expected distributions of convergence, shear, and source position, we obtained the population-averaged retention factor, which quantifies the expected degradation in the Bayes factor across realistic lens populations. The results, presented in Figure~\ref{fig:PMX-blu_ratio}, demonstrate that the use of an incomplete model leads to a systematic underestimation of the Bayes factor in the high lens-mass regime.

These findings underscore the importance of including galaxy effects in the models of lensed waveforms when searching for microlensed GW signals. Neglecting such effects may result in reduced sensitivity to microlensing events in current and future GW datasets. In addition, this mismatch could weaken constraints on the compact dark matter fraction $f_\mathrm{DM}$ derived using PML-only searches (e.g.,~\citep{Basak:2021ten, LIGOScientific:2023bwz}). Quantifying the change requires recomputing the population-averaged detection efficiency using PMX signals. If we assume a spatial distribution of the microlenses following the stellar luminosity (i.e., assuming that these microlenses are astrophysical black holes, not PBHs), then a larger fraction of them will reside in the deeper regions of the galactic potential, and hence the lensing contribution due to the galaxy will be larger. Hence our estimate on the retention factor is likely to be conservative. 

Note that our current study suffers from some limitations and should be treated only as a proof-of-concept study. First, the macromodel of the host galaxy was assumed to be modeled by SIS, which, while analytically convenient, does not fully capture the diversity of galactic potentials. In particular, elliptical galaxy profiles may introduce different distributions of shear and convergence, thereby altering the PMX parameter space. Second, the galaxy parameters are fixed to values representative of the Milky Way. A more realistic treatment will draw these parameters from a distribution of galaxy lenses obtained from a combination of galaxy surveys and theoretical modeling. Finally, our analysis assumes a high-SNR approximation to the Bayes factor, which may not hold for all observed events.

A more rigorous treatment would require assessing model evidence within a full Bayesian framework, rather than relying on the large-SNR approximation adopted here. It also involves using more realistic models for the background galaxy and its parameters. Since a full Bayesian analysis using more complex lens models is computationally prohibitive using current techniques, we defer such an analysis for future work. The surrogate modeling techniques introduced in \citep{Deka:2025vzx} may provide a practical resolution to this issue. We also intend to rederive the upper limits on $f_\mathrm{DM}$ in the future.


\begin{acknowledgements}
    We thank the members of the Astrophysics \& Relativity group at ICTS for their valuable inputs. We thank the anonymous referee for pointing out a normalization error in the earlier version of Figure 8. S.B. would like to thank ICTS-TIFR for hospitality throughout the course of this study. We acknowledge the support of the Department of Atomic Energy, Government of India, under Project No. RTI4001. The numerical calculations were performed using the Alice computing cluster at ICTS-TIFR. This material is based upon work supported by NSF's LIGO Laboratory, which is a major facility fully funded by the National Science Foundation.
\end{acknowledgements}


\appendix

\section{Derivation of the retention factor}\label{app:loss_factor_derivaiton}

\begin{figure*}[t]
    \centering
    \includegraphics[width=\textwidth]{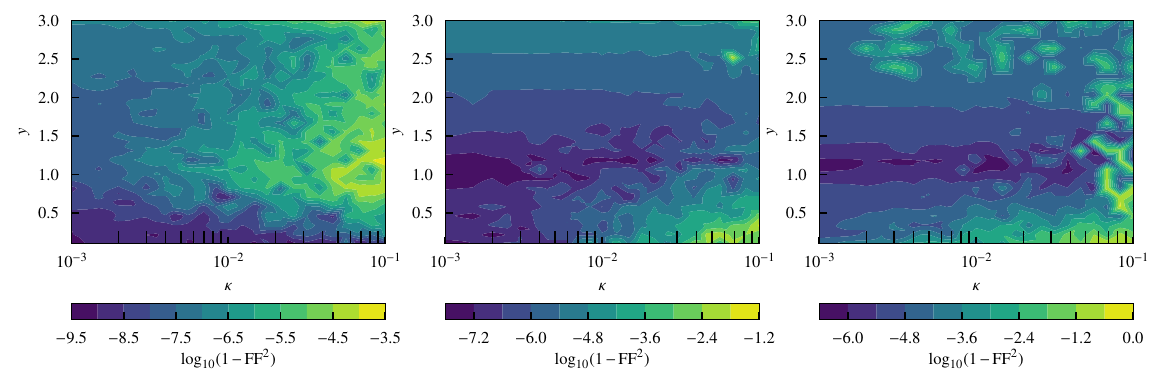}
    \caption{Fitting factor between the PMX-lensed signal and the PML template in the $\kappa-y$ plane for a PMX lens for redshifted lens masses $M_{\ell z} = 10^2, 10^3, 10^4 M_\odot$ (from left to right). The shear angle values at each point on the plane are randomly drawn from a uniform distribution over $[0, 2\pi)$. This random sampling introduces the small-scale nonsmoothness visible in the plot. The BBH source is nonspinning, with chirp mass $\mathcal{M}_c=40 M_\odot$ and symmetric mass ratio $\eta=0.2$. Colors indicate $\log_{10}(1-\mathrm{FF}^2)$, where smaller values indicate better agreement between the PML and PMX waveforms.}
    \label{fig:FF_plots}
\end{figure*}

The true signal is assumed to be lensed by a microlens in an external potential, so that it is well described by a PMX lens. The parameters describing the lensed signal are $\paramsTrue = \{\vec{\vartheta}^{\rm tr},\vec{\Lambda}_\ext^{\rm tr}\}$, where $\vec{\vartheta}^{\rm tr}$ refers to the true parameters of the binary source and $\vec{\Lambda}_\ext^{\rm tr}$ are the true lens parameters. 

Under the hypothesis that the waveform is being microlensed by a $\ext$ lens, the zero-noise likelihood\footnote{This is equivalent to retaining the parameter-dependent part of the likelihood obtained from the log-likelihood averaged over many realizations of zero-mean Gaussian noise.} maximized over the distance to the source can be written as
\begin{equation}\label{eq:likelihood_ext}
  p(d\mid\vec\theta_\ext, \hyp{\rm PMX}) = \exp\left[-\frac{\rho^2}{2} ~\left(1 - \mathcal{M}^2(\paramsTrue,\vec\theta_\ext)\right)\right]~,
\end{equation}
where $\vec\theta_\ext\equiv\{\vec{\vartheta}, \vec{\Lambda}_{\ext}\}$ is the set of source and lens parameters describing the PMX template waveform, $\rho$ is the optimal SNR  of the true signal, while the \emph{match}, $\mathcal{M}$, between two waveforms is defined as~\citep{Owen:1995tm},
\begin{equation}\label{eq:match_definition}
  \mathcal{M}(\vec{\theta},\vec{\theta}\,')\equiv\max_{\Delta t_c}\left|4\int_{f_{\rm low}}^{f_{\rm upp}}\frac{\hat{h}^{*}(f;\vec{\theta})\,\hat{h}(f;\vec{\theta}\,')\,e^{2\pi i f\Delta t_c}}{S_n(f)}\,df\right|, 
\end{equation}
where $S_n(f)$ is the one-sided PSD of the detector noise, an asterisk denotes complex conjugation, and $\hat{h}(f;\vec\theta)$ denotes the normalized frequency-domain GW signal with parameters $\vec \theta$, normalized such that
\begin{equation}\label{eq:normalized_waveform}
    4\int_{f_{\rm low}}^{f_{\rm upp}}\frac{\left|\hat{h}(f;\vec{\theta})\right|^2}{S_n(f)}\,df=1.
\end{equation}
The maximization over $\Delta t_c$ accounts for a relative shift in coalescence time, while the absolute value maximizes over an overall phase shift. We adopt the frequency range $f_{\rm low}=18$ Hz and $f_{\rm upp}= 1024$ Hz throughout this study.

In the same way, in a search for the same signal using the unlensed waveform model, we get the likelihood to be
\begin{equation}\label{eq:likelihood_ul}
  p(d\mid\vec{\vartheta}, \hyp{\rm U}) = \exp\left[-\frac{\rho^2}{2}\left(1 - \mathcal{M}^2(\paramsTrue,\vec{\vartheta})\right)\right]~.
\end{equation}

The corresponding evidences, assuming flat priors on all parameters, are
\begin{subequations}\label{eq:evidences}
  \begin{align}
    p(d\mid \hyp{\rm PMX}) &= \int \, d\vec{\theta}_\mathrm{PMX} \, p(d\mid\vec\theta_\ext, \hyp{\rm PMX}) ,\\
    p(d\mid \hyp{\rm U}) &= \int \, d\vec{\vartheta}\, p(d\mid\vec{\vartheta}, \hyp{\rm U})~,
  \end{align}
\end{subequations}
respectively. In the large-SNR limit, the Bayes factor between the ``PMX-lensed'' and ``unlensed'' hypotheses can be approximated as~\citep{Cornish:2011ys}
\begin{equation}
\label{eq:bayes_factor_ext_U}
  \mathcal{B}_{\rm U}^{\rm PMX} = \frac{p(d\mid \hyp{\rm PMX})}{p(d\mid \hyp{\rm U})}
  \approx \frac{\exp\left[\frac{-\rho^2}{2}\left(1 - \mathrm{FF}^2(\vec\theta_{\rm PMX})\right)\right]}{\exp\left[\frac{-\rho^2}{2}\left(1 - \mathrm{FF}^2(\vec{\vartheta})\right)\right]}, 
\end{equation}
where $\mathrm{FF}(\vec\theta_{\rm PMX})$  and $\mathrm{FF}(\vec{\vartheta})$ denote the \emph{fitting factor} of the normalized PMX template waveform with the normalized unlensed template waveform, respectively, with the PMX-lensed signal. The log Bayes factor is 
\begin{align}
 \ln\mathcal{B}_{\rm U}^{\rm PMX} \approx \frac{\rho^2}{2}\left[\mathrm{FF}^2(\vec\theta_{\rm PMX}) -  \mathrm{FF}^2(\vec{\vartheta})\right] \approx \frac{\rho^2}{2}\left[1 - \mathrm{FF}^2(\vec{\vartheta}) \right],
\end{align}
where, in the second step, we use the fact that $\mathrm{FF}(\vec\theta_{\rm PMX}) \approx 1$. {Note that we retain only the leading maximum-likelihood contribution to the evidence ratio and neglect model-dependent prior- and posterior-volume (Occam) terms.~\citep{Cornish:2011ys}}

Current searches use an isolated PML model to search for the PMX-lensed signal. Following the steps above, we can derive the corresponding log Bayes factor as
\begin{align}
 \ln\mathcal{B}_{\rm U}^{\rm PML} \approx \frac{\rho^2}{2}\left[\mathrm{FF}^2(\vec\theta_{\rm PML}) -  \mathrm{FF}^2(\vec{\vartheta}) \right],
\end{align}
where $\mathrm{FF}(\vec\theta_{\rm PML})$ is the fitting factor of the PML template toward the PMX-lensed signal. The suboptimal search using the isolated PML templates should produce this Bayes factor. Now, we define the optimality of the search as 
\begin{align}\label{eq:loss_factor}
  \frac{\mathcal{B}_{\rm U}^{\,\rm PML}}{\mathcal{B}_{\rm U}^{\rm PMX}} = \exp\left[\frac{-\rho^2}{2}\left(1 - \mathrm{FF}^2(\vec{\theta}_{\rm PML})\right)\right]~.
\end{align}
If this ratio is close to 1, current searches using PML templates are already close to being optimal. If the ratio is significantly smaller than 1, that means that the PML searches are not optimal for detecting PMX lensed signals.  

Figure~\ref{fig:FF_plots} shows the $\mathrm{FF}(\vec{\theta}_{\rm PML})$ in the $\kappa-y$ plane for three different values of redshifted lens mass. As is evident from the figure, high-mass lenses yield larger values of $(1-\mathrm{FF}^2)$, corresponding to a poorer overlap between the PML template and the PMX-lensed signal. This trend is consistent with the results shown in Figure~\ref{fig:PMX-blu_ratio}, where the retention factor decreases with increasing lens mass.

\section{Validity of the lens model with constant \texorpdfstring{$\kappa$}{kappa} and \texorpdfstring{$\gamma$}{gamma}}\label{app:model_validity}

The characteristic distance scale over which wave effects are significant for microlensing is set by the Fresnel radius, which is given by~\citep{Macquart:2004sh},
\begin{align}\label{eq:r_fresnel}
  r_F \equiv \sqrt{\frac{1}{2\pi f (1+z_\ell)}\frac{\dl\dls}{\ds}} = \frac{\xi_0}{\sqrt{w}}~,
\end{align}
where the quantities are defined in Sec.~\ref{sec:lensing_diffraction}.

\begin{figure}[htbp]
  \centering
  \includegraphics{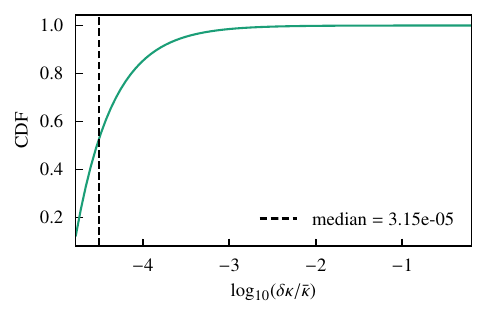}
  \caption{Cumulative distribution of the variation of $\kappa$ across the host galaxy measured within regions of size equal to the largest Fresnel radius of the microlens.}
  \label{fig:rF_variation}
\end{figure}

To assess the variation of $\kappa$ (and in turn $\vec\gamma$) in the Fresnel zone of the microlens, we partition the radial coordinate $\xi_\bullet$ on the lens plane (which sets the location of the microlens in the galaxy) into a grid of cells of size $r_F$. We adopt the largest Fresnel radius considered in this study, making this a conservative test. Within each cell, $\kappa$ (and consequently $\vec{\gamma}$) may be treated as effectively constant.  Specifically, the fractional variation of $\kappa$ across a Fresnel-scale cell, $\delta\kappa/\bar{\kappa}$, is typically below $10^{-4}$ (Figure~\ref{fig:rF_variation}). Here $\bar{\kappa}$ denotes the cell averaged value of $\kappa$ and $\delta\kappa\equiv\textrm{max}(\kappa)-\textrm{min}(\kappa)$. Although $\delta\kappa/\bar{\kappa}$ can approach unity for very small values of $\xi_\bullet$, the expected number of microlenses in this region is very small (see Figure~\ref{fig:microlens_spatial_dist}). We therefore treat $\kappa$ and $\gamma$ as locally constant close to each microlens, as assumed in deriving the PMX lensing potential in Equation~\eqref{eq:lensing_potential_PMX}.


\bibliography{references}{}
\bibliographystyle{aasjournalv7}

\end{document}